\documentclass[lettersize,journal]{IEEEtran}
\usepackage{amsmath,amsfonts}
\usepackage{array}
\usepackage[caption=false,font=normalsize,labelfont=sf,textfont=sf]{subfig}
\usepackage{textcomp}
\usepackage{stfloats}
\usepackage{url}
\usepackage{verbatim}
\usepackage{cite}
\usepackage{graphicx}
\usepackage{hyperref}
\usepackage{multirow}
\usepackage{booktabs}
\usepackage{enumitem}
\usepackage{color}
\usepackage{subfig}
\usepackage{float}
\usepackage{wrapfig}
\usepackage{tabularx}
\usepackage{mathbbol}
\usepackage{amsmath}
\usepackage{amssymb}
\usepackage[table]{xcolor}
\usepackage{colortbl}
\usepackage{stfloats}
\usepackage{float}
\usepackage{color}
\usepackage[ruled]{algorithm2e}
\usepackage[shortcuts]{extdash}

\usepackage{newfloat}
\usepackage{listings}

\begin{document}
\title{DH-Mamba: Exploring Dual-domain Hierarchical State Space Models for MRI Reconstruction}
\author{Yucong Meng, Zhiwei Yang, Kexue Fu, Zhijian Song, Yonghong Shi
\thanks{Y.Meng, Y.Shi, and Z.Song are with Digital Medical Research Center, School of Basic Medical Science, Fudan University, Shanghai 200032, China, and also with the Shanghai Key Laboratory of Medical Image Computing and Computer Assisted Intervention, Shanghai 200032, China. Z.Yang is with Academy of Engineering and Technology, Fudan University, and also with the Shanghai Key Laboratory of Medical Image Computing and Computer Assisted Intervention, Shanghai 200032, China. K.Fu is with Shandong Computer Science Center (National Supercomputer Center in Jinan), China.}
\thanks{Corresponding authors: Zhijian Song, Yonghong Shi. (email: zjsong@fudan.edu.cn; yonghong.shi@fudan.edu.cn)}
}

\markboth{Journal of \LaTeX\ Class Files,~Vol.~14, No.~8, August~2021}%
{Shell \MakeLowercase{\textit{et al.}}: A Sample Article Using IEEEtran.cls for IEEE Journals}

\IEEEpubid{0000--0000/00\$00.00~\copyright~2021 IEEE}

\maketitle

\begin{abstract}
The accelerated MRI reconstruction poses a challenging ill-posed inverse problem due to the significant undersampling in k-space. Deep neural networks, such as CNNs and ViTs, have shown substantial performance improvements for this task while encountering the dilemma between global receptive fields and efficient computation. To this end, this paper explores selective state space models (Mamba), a new paradigm for long-range dependency modeling with linear complexity, for efficient and effective MRI reconstruction. However, directly applying Mamba to MRI reconstruction faces three significant issues: (1) Mamba typically flattens 2D images into distinct 1D sequences along rows and columns, disrupting k-space's unique spectrum and leaving its potential in k-space learning unexplored. (2) Existing approaches adopt multi-directional lengthy scanning to unfold images at the pixel level, leading to long-range forgetting and high computational burden. 
(3) Mamba struggles with spatially-varying contents, resulting in limited diversity of local representations. To address these, we propose a dual-domain hierarchical Mamba for MRI reconstruction from the following perspectives: (1) We pioneer vision Mamba in k-space learning. A circular scanning is customized for spectrum unfolding, benefiting the global modeling of k-space. (2) We propose a hierarchical Mamba with an efficient scanning strategy in both image and k-space domains. It mitigates long-range forgetting and achieves a better trade-off between efficiency and performance. (3) We develop a local diversity enhancement module to improve the spatially-varying representation of Mamba. Extensive experiments are conducted on three public datasets for MRI reconstruction under various undersampling patterns. Comprehensive results demonstrate that our method significantly outperforms state-of-the-art methods with lower computational cost. Code will be available in \href{https://github.com/XiaoMengLiLiLi/DH-Mamba}{https://github.com/XiaoMengLiLiLi/DH-Mamba}.
\end{abstract}

\section{Introduction}
\IEEEPARstart{M}{agnetic} Resonance Imaging (MRI) is a crucial clinical tool due to its non-radiation, high resolution, and superior contrast~\cite{r1,katti2011magnetic}. However, its long scan time causes patient discomfort and image blurring, limiting its application in dynamic imaging and real-time diagnosis. To alleviate this bottleneck, researchers have developed two main MRI acceleration paradigms. (1) Parallel Imaging (PI) \cite{r43}, reducing the scan time by using multiple receiver coils to acquire data simultaneously. However, the acceleration factor of PI is limited by the number and arrangement of the receiver coils, and it also increases the manufacturing cost of the MRI scanner. (2) Compressed Sensing (CS) \cite{r2}, bypassing the Nyquist-Shannon sampling criteria with more aggressive undersampling. As a more economical type with less dependence on hardware, CS has received extensive interest. In this paper, we focus on compressed sensing based MRI reconstruction (CS-MRI). The main challenge is to find an algorithm to accurately reconstruct high-quality MRI images without artifacts from highly undersampled k-space data.

\begin{figure}[!t]
\centerline{\includegraphics[width=1.0\linewidth]{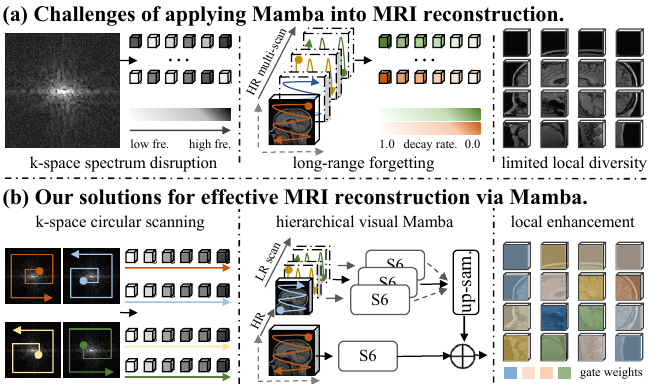}}
\caption{Our main idea. (a) Applying vanilla Mamba into MRI reconstruction faces three challenges: the disruption of k-space spectrum, the long-range forgetting, and the lack of local diversity. (b) To address these, we pioneer a dual-domain hierarchical Mamba for MRI reconstruction. It tackles above issues through circular scanning in k-space, hierarchical Mamba operation in both image and k-space domains, and local diversity enhancement.}
\label{Fig.1}
\end{figure} 

In recent years, with the rise of deep learning, many methods have adopted different Convolutional Neural Networks (CNNs) for MRI reconstruction~\cite{r3,r4,CAMPNet}. Benefiting from CNNs' powerful nonlinear representations, these approaches have surpassed traditional MRI reconstruction methods. However, CNNs often fail to model long-range dependencies, and the limited Effective Receptive Fields (ERF) trapped it in a bottleneck~\cite{r42}. To overcome these limitations, Vision Transformers (ViTs) have been applied to MRI reconstruction~\cite{r5,r6,10559239}. Due to the unique self-attention mechanism, ViTs hold increased ERF. It allows the network to gather information from a broader area and identify higher-level patterns within the images. Based on these compelling properties, ViTs  \IEEEpubidadjcol have been the mainstream for MRI reconstruction, demonstrating substantial performance improvements~\cite{FPS,r27,10606502,9393325}.

However, the self-attention mechanism of ViTs has an inherent problem where computational complexity grows quadratically with input size~\cite{r7,r8,r40,r41}. This limits ViTs' ability to efficiently model images at high spatial resolutions, hindering the performance of MRI reconstruction. Although some approaches utilized shifted window mechanisms~\cite{r9} to alleviate this problem, they essentially sacrifice the global receptive field and fail to fundamentally balance the trade-off between global effectiveness and computational efficiency.

Considering the limitations of CNN-based and ViT-based methods, selective state space models (Mamba)~\cite{r10,10857393,10770251,10794530} has emerged as a novel paradigm, offering new insights for MRI reconstruction with an optimal balance between performance and efficiency. By leveraging input-dependent state space models to compress global context, Mamba achieves linear complexity while preserving a global receptive field. This enables Mamba-based networks to unfold images at pixel-level and construct ultra-long sequences. Therefore, they can capture fine-grained features and shed new light on enhancing MRI reconstruction with less computational cost~\cite{r11,r12}.

However, applying vanilla Mamba to MRI reconstruction presents three key challenges, as shown in Fig.~\ref{Fig.1} (a). Specifically, (1) Vision Mamba’s row-wise and column-wise scanning strategy is designed for the natural images, while k-space's unique frequency arrangement places low frequencies at the center and high frequencies around the periphery. Directly applying vanilla scanning to k-space spectrum leads to the disruption of frequency structures. (2) Existing approaches’ token processing leads to the decay of the previous hidden state, resulting in long-range forgetting and high computational cost. It loses high-level features relevant to the query pixel, limiting the model's ability to capture global dependencies in ultra-long sequences as demonstrated in~\cite{yuan2024remamba,zhang2024hrvmamba}. (3) Mamba aggregates features via pixel-wise linear scaling, overlooking spatially-varying representations. This results in limited local diversity and hinders accurate MRI reconstruction.

To address these issues, we propose Dual-domain Hierarchical Mamba (DH-Mamba), which carefully adapts Mamba for both image and k-space domains for the first time. It provides an effective and efficient paradigm for MRI reconstruction, as illustrated in Fig.\ref{Fig.1} (b). Specifically, (1) To preserve the frequency structure of k-space, a circular scanning scheme is deigned. It reorganizes the frequencies from low to high, benefiting the global modeling of k-space. (2) To mitigate the long-range forgetting and reduce the computational cost, a hierarchical Mamba structure is proposed in both image and k-space domains. We categorize the scanning directions into two groups: one retains the high resolution and the other three process on the downsampled feature maps. It efficiently captures both fine and coarse-level features with less computational burden. (3) To introduce local diversity during feature propagation, a local enhancement module is developed. It multiplies feature maps by a learnable pixel-wise gating mask and enables selective contribution of different local features, enhancing spatially varying feature representation.

Our main contributions are summarized as follows:

\begin{itemize}
    \item The potential of Mamba for MRI reconstruction in both image and k-space domains is explored for the first time, providing an effective and efficient solution to this challenging task.
    \item A circular scanning strategy is designed to effectively organize the relationship between different frequencies, preserving the unique structure of the k-space.
    \item A hierarchical Mamba structure is proposed to mitigate the long-range forgetting and information redundancy while minimizing the computational complexity. 
    \item A local enhancement module is developed to enhance the local diversity of feature representations, resulting in more accurate and reliable MRI reconstruction.
\end{itemize}

\section{RELATED WORK}
\subsection{MRI Reconstruction}
Magnetic Resonance Imaging (MRI), as an advanced medical diagnostic tool, can provide high-quality anatomical information. However, traditional MRI requires densely sampled k-space data, leading to long scan time, patient discomfort, and limited clinical throughput. Currently, MRI acceleration is achieved through two technical approaches: Parallel Imaging (PI)\cite{r43} and Compressed Sensing (CS)\cite{r2}. PI speeds up scanning by using multiple receiver coils. However, its acceleration rate is limited by the number and arrangement of coils, and it increases hardware cost. In contrast, CS bypasses the Nyquist-Shannon sampling theorem, achieving a high acceleration factor through undersampling strategies and reconstruction algorithms. CS relies less on hardware and is a more cost-effective solution, making it an attractive option. However, traditional CS-MRI suffers from high complexity and severe artifacts. Recently, deep learning shown great potential in MRI reconstruction. Related methods are mainly divided into two categories, i.e., CNNs based and ViTs based approaches.

\subsubsection{CNNs Based Methods} 
The rise of deep learning has significantly advanced fast MRI, with numerous CNN-based methods achieving remarkable success. For example, Wang et al.~\cite{r16} first applied deep learning to MRI reconstruction by proposing an offline CNN method. D5C5~\cite{r17D5C5} reconstructed dynamic sequences of 2D cardiac MRI images via a deep cascade of CNNs. DuDoRNet \cite{r21} incorporated T1 priors for simultaneous k-space and image restoration. DCRCN \cite{r19DCRCN} presented a novel MRI recovery framework via densely connected residual blocks. Dual-OctConv \cite{r20} learned multi-scale spatial-frequency features from both real and imaginary components for parallel MRI reconstruction. To sum up, CNN-MRI leverages CNNs' nonlinearity to learn intricate patterns and features from extensive datasets, enabling them to effectively restore images from low-quality inputs. However, CNNs fail to learn long-range dependencies due to the limited receptive field of convolutional layers, restricting the further development of such methods\cite{r22,r23,sahu2021survey,khan2023survey}.

\subsubsection{ViTs Based Methods} 
To overcome the above limitations of CNNS, Vision Transformers (ViTs) ~\cite{IGKR,MoRe,UNIA} have been introduced for MRI reconstruction. ViTs leverage self-attention mechanism to achieve global interaction, thus have increased receptive fields and improved reconstruction performance. Lin et al.\cite{r6} demonstrated that a ViT tailored for image reconstruction yields on par reconstruction accuracy with UNet while enjoying higher throughput and less memory consumption. Korkmaz et al.\cite{r24} introduced SLATER, achieving unsupervised MRI reconstruction based on zero-shot learned adversarial transfomers. SwinMR \cite{r25} introduced swin-transformer to MRI reconstruction, realizing effective fast parallel imaging. FMTNet \cite{r26} proposed to separate different frequencies and model them individually, reconstructing MRI images with clear structure. ReconFormer \cite{r27} incorporated a local pyramid and global columnar ViT to learn multi-scale feature at any stage, enabling enhanced reconstruction performance. However, the self attention calculation results in ViTs' quadratic computational complexity with respect to sequence length. This inherent limitation confines ViTs-based methods to processing images in coarse patches, significantly restricting their ability to capture fine-grained features, which are crucial for MRI reconstruction. 

\subsection{State Space Models and Mamba}
State Space Models (SSMs) \cite{r46,r29,r30} provide a framework for modeling system evolution over time, using state and observation equations. They are widely applied in various tasks, such as control theory and signal processing. Recently, Mamba \cite{r10}, as one of the most prominent SSMs, has received increasing attention. Mamba combined selective scanning (S6) with data-dependent parameters, achieving global receptive fields while enjoying linear complexity~\cite{xu2024visual,rahman2024mamba,liu2024swin,liu2024vision,ibrahim2025survey,qiao2024vl}. 

Due to Mamba's superior ability to capture global information, several studies have leveraged it for image restoration tasks. 
MambaIR \cite{r31} introduced local convolution to adapt Mamba to low-level vision and proposed an image recovery baseline. VMambaIR \cite{r32} designed six-directions omni selective scan and incorporated it into a UNet architecture for various image restoration tasks. Unlike previous methods that solely consider image domain's properties such as locality and continuity, we tailor a dual-domain Mamba structure for both k-space and image processing, for MRI reconstruction. It features the innovations of effective k-space learning, hierarchical SSMs modeling, and local diversity preserving.

For frequency domain explorations, FreqMamba \cite{zhen2024freqmamba} realized that the Fourier transform can disentangle image content and raindrop, and proposed to decompose the image into sub-bands of different frequencies to allow 2D scanning from the frequency dimension for image deraining. FAHM \cite{10877784} demonstrated that high and low frequencies contain different image contents, and introduced a frequency embedding module to extract various frequency components for hyperspectral image classification. Unlike existing approaches that consider the relationship between the image contents and the Fourier domain for natural image tasks, we focus on the unique properties of the k-space raw data and customize a circular scanning strategy for effective k-space learning. 

For MRI reconstruction, MambaMIR \cite{MambaMIR} proposed an arbitrary-mask mechanism, which adapts Mamba to fast MRI reconstruction and uncertainty estimation. However, it didn't take k-space information into account, achieving suboptimal results. MMR-Mamba \cite{MMRMamba} used Mamba to integrate multi-modality features for MRI reconstruction, with the help of fully-sampled auxiliary modality. However, the previous studies are limited to applying general Mamba to MRI reconstruction, and there are few attempts to design a customized Mamba structure for this task. Research on Mamba in k-space modeling remains scarce, leaving effective and efficient MRI reconstruction being largely unexplored. Differently from existing works, we pioneer to explore Mamba in k-space learning and customize a dual-domain hierarchical Mamba, achieving significant performance improvement without the need of a fully-sampled auxiliary modality.

\section{METHODOLOGY}
\begin{figure*}[!t]
\centerline{\includegraphics[width=1.0\linewidth]{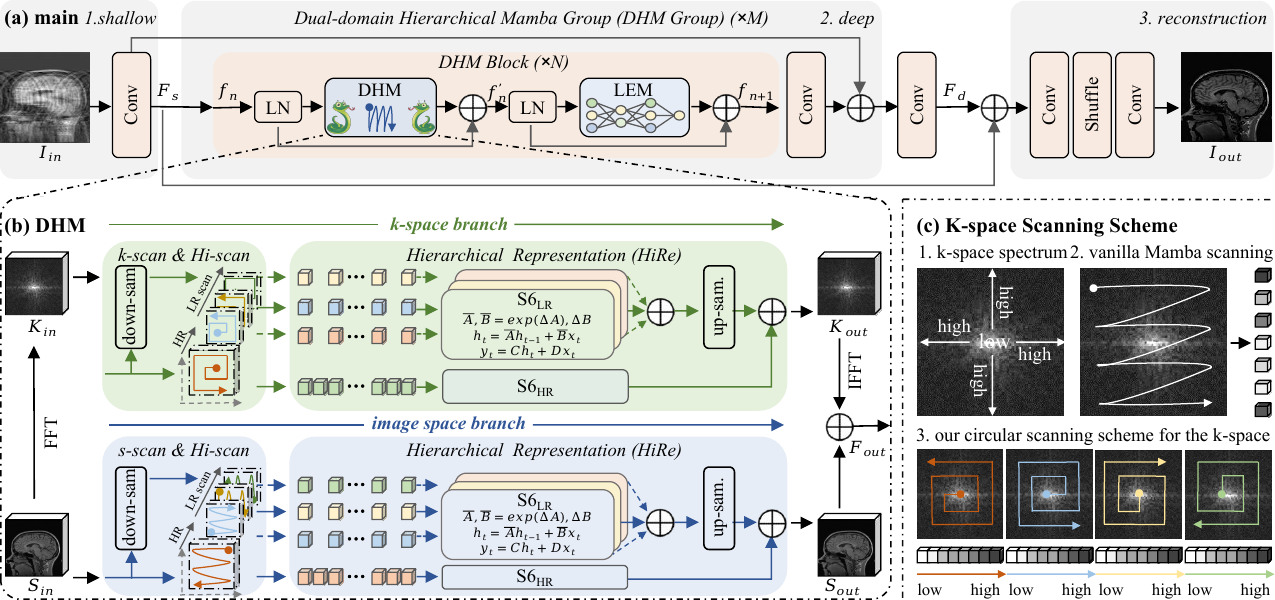}}
\caption{(a) The overall architecture of the proposed DH-Mamba, which can be divided into three stages, i.e., shallow extraction, deep extraction, and high quality reconstruction. Given the low-quality MRI image $I_{in}$ as input, we first obtain the shallow features $F_{s}$. Then, we send $F_{s}$ into the network backbone, i.e., stacked DHM Groups, to extract deep features $F_{d}$. The core design of DHM Group is stacked DHM Blocks, which consists of DHM and LEM. Finally, the fused $F_{s}$ and $F_{d}$ is send into reconstruction head to obtain the high-quality output $I_{out}$. (b) The proposed DHM consists of two branches, i.e., k-space branch and image space branch, processing in the k-space and image domains, respectively . (c) The motivation of our k-space scanning. General scanning disrupts the k-space's structure, then we customize the circular scanning scheme to rearrange frequencies.}
\vspace{-0.2 cm}
\label{Fig.2}
\end{figure*}

\begin{figure}[!t]
\centerline{\includegraphics[width=\columnwidth]{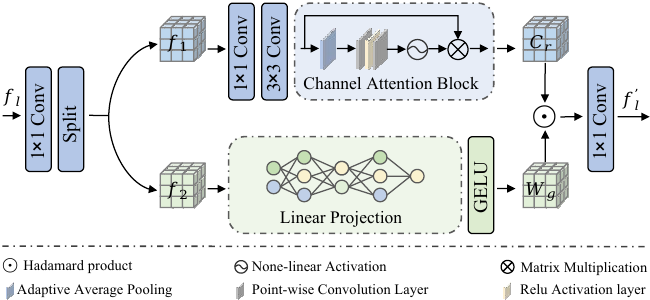}}
\caption{Architecture of the proposed local enhancement module (LEM). Given the input features, after expending the channels, they are split into two parts along the channel dimension and processed by two branches, respectively. Then, the resulting outputs are combined via Hadamard product and convolution to obtain the final output.}
\label{Fig.3}
\vspace{-0.5 cm}
\end{figure}

\subsection{Preliminaries}
\subsubsection{MRI Reconstruction} Let $K$ represent the fully sampled k-space acquired from the MRI scanner, accelerated MRI employs undersampling to acquire a reduced set of k-space, i.e., $K_{s}$. Here, we simulate this process by the element-wise multiplication ($\otimes$) of $K$ with a two-dimensional mask $M$:
\begin{equation} K_{s}=M\otimes K.
\label{eq1}\end{equation}

Correspondingly, low-quality MRI image $I_{s}$ can be obtained, i.e., $I_{s} = IFFT(K_{s})$, where $IFFT(\cdot)$ denotes Inverse Fast Fourier Transform. To recover high-quality MRI image $\widehat{{I}}$, deep learning methods typically leverage extensive training data to establish a mapping relationship between $\widehat{{I}}$ and the sampled data $(I_{s}, K_{s})$, which can be formulated as:
\begin{equation}\widehat{{I}}=f_{\theta}(I_{s},K_{s}),
\label{eq2}\end{equation}
where $\theta$ is the parameters set of the deep neural network.

\subsubsection{Selective State Space mechanism of Mamba (S6)}
As one of the most prominent State Space Models (SSMs), Mamba enhances performance through its core design, i.e., the Selective State Space mechanism (S6). Essentially, classical SSM is a framework of linear Ordinary Differential Equations (ODEs), which transforms an input sequence $x(t)\in \mathbb{R}^{L}$ to a latent representation $h(t)\in \mathbb{R}^{H}$ and subsequently predicts an output sequence $y(t)\in \mathbb{R}^{L}$. Mathematically, the process of conventional SSM can be formulated as follows:
\begin{equation}
{h}'(t)= \textbf{A}h(t)+\textbf{B}x(t), y(t)=\textbf{C}h(t)+\textbf{D}x(t),
\label{eq3}\end{equation}
where $H$ denotes the size of the state, the parameters $\textbf{A}\in \mathbb{R}^{H\times H}$, $\textbf{B}\in \mathbb{R}^{H\times 1}$, $\textbf{C}\in \mathbb{R}^{1\times H}$ are associated with $H$, and $\textbf{D}\in \mathbb{R}^{1}$ represents the skip connection. As a continuous-time system, SSM presents challenges for direct integration into deep learning. To address this, previous studies achieved discretization via the zero-order hold method. Specifically, they use a timescale parameter $\triangle$ to discretize the continuous parameters $\textbf{A}$ and $\textbf{B}$, obtaining $\overline{\textbf{A}}$ and $\overline{\textbf{B}}$ as follows:
\begin{equation}
\overline{\textbf{A}}=exp(\triangle \textbf{A}), \overline{\textbf{B}}=(\triangle A)^{-1}(exp(A)-\textbf{I}) \triangle \textbf{B},
\label{eq4}\end{equation}
where \textbf{I} denotes the identity matrix. This results in the discretized form of Eq.(\ref{eq3}):
\begin{equation}
h(k)= \overline{\textbf{A}}h(k-1)+\overline{\textbf{B}}x(k), y(k)=\textbf{C}h(k)+\textbf{D}x(k).
\label{eq5}\end{equation}
After that, the process of Eq.(\ref{eq5}) can be implemented in convolutional manner as follows:
\begin{equation}
y=x\odot \overline{\textbf{K}}, \overline{\textbf{K}}=(\textbf{C}\overline{\textbf{B}}, \textbf{C}\overline{\textbf{AB}},\dots,\textbf{C}\overline{\textbf{A}}^{L-1}\overline{\textbf{B}}),
\label{eq6}\end{equation}
where $\overline{\textbf{K}}$ is the convolution kernel.
More recently, Mamba proposed S6 to further render $\overline{\textbf{B}}$, $\overline{\textbf{C}}$ and $\triangle$ to be input-dependent, allowing for a dynamic feature representation as:
\begin{equation}
y=x\odot \overline{\textbf{K}}, \overline{\textbf{K}}=(\textbf{C}_{L}\overline{\textbf{B}}_{L}, \textbf{C}_{L}\overline{\textbf{A}}_{L-1}\overline{\textbf{B}}_{L-1},\dots,\textbf{C}_{L}\prod_{i=1}^{L-1} \overline{\textbf{A}_{i}\textbf{B}_{1}}).
\label{eq7}\end{equation}

\subsection{Overall Pipeline}
Following previous work ~\cite{r11}, DH-Mamba is built up with three stages, i.e., shallow feature extraction implemented by convolution layers, deep feature acquisition via stacked Dual-domain Hierarchical Mamba Groups (DHM Groups), and high-quality MRI reconstruction through convolution and pixel-shuffle layers, as shown in Fig.\ref{Fig.2} (a). 
\subsubsection{Shallow Extraction}
Given low-quality MRI image $I_{in}\in \mathbb{R}^{H \times W \times 2}$ with spatial resolution $H \times W$, we first employ a $3 \times 3$ convolution $conv (\cdot)$ to generate shallow features $F_{s}\in \mathbb{R}^{H \times W \times C}$, where $C$ denotes the channel dimension:
\begin{equation}
F_{s}=conv(I_{in}).
\label{eq8}\end{equation}
\subsubsection{Deep Extraction}
Subsequently, $F_{s}$ undergoes a deep feature extractor consisting of $M$ DHM Groups, with each DHM Group containing $N$ DHM Blocks. Specifically, given $f_{n}\in \mathbb{R}^{H \times W \times C}$, $n \in \left \{ 1,2,\dots ,N \right \} $ as input, the processing of $n\-/th$ DHM Block can be formulated as follows:
\begin{equation}
\begin{split}
f^{'}_{n}&=\alpha \cdot f_{n}+DHM(LN(f_{n})) \\ f_{n+1}&=\beta \cdot f^{'}_{n}+LEM(LN(f^{'}_{n})),
\label{eq9}
\end{split}
\end{equation}
where $LN(\cdot)$ is the layer normalization, $\alpha \in \mathbb{R}^{C}$ and $\beta \in \mathbb{R}^{C}$ are learnable factors to control the skip connection. The Dual-domain Hierarchical Mamba (DHM) and Local Enhancement Module (LEM) are two core designs of our DHM Block. 

When $n=N$, the obtained $f_{N+1}$ is taken as the input of DHM Group. And the process of the $m\-/th$ DHM Group, $m \in \left \{ 1,2,\dots ,M \right \}$, can be written as follows:
\begin{equation}
f_{m+1} = f_{m} + conv(f_{m}).
\label{eq10}
\end{equation}
When $m=M$, we take the obtained $f_{M+1}$ as the output of deep extractor and further use a convolution layer to obtain the deep feature $F_{d}$. This process can be written as follows:
\begin{equation}
F_{d}=conv(f_{M+1}).
\label{eq11}
\end{equation} 
\subsubsection{High Quality Reconstruction}
Finally, we perform an element-wise addition of $F_{s}$ and $F_{d}$. The resulted features are fed into a reconstruction head $recon (\cdot)$, which consists of convolution and pixel-shuffle layers to reconstruct $I_{out}$ as:
\begin{equation}
I_{out}=recon(F_{s}+F_{d}).
\label{eq12}
\end{equation}
Overall, the above reconstruction process can be summarized as: $I_{out}=\mathcal{N}(I_{in})$, where $\mathcal{N}(\cdot )$ is the overall network and is trained by minimizing the following loss function:
\begin{equation}
    \mathcal{L} =\left \| I_{out}- I_{gt}  \right \| _{1},
\label{eq13}
\end{equation}
where $I_{gt}$ is the ground truth, and $\left \| \cdot  \right \| _{1} $ is the $L_{1}$-norm.

\subsection{Dual-domain Hierarchical Mamba (DHM)} 
As elaborated above, DHM serves as the core module of DH-Mamba, which consists of two branches as shown in Fig.\ref{Fig.2} (b). To be specific, the k-space branch processes in the k-space domain, and the image space branch operates on the spatial domain. Specifically, given the input feature $S_{in}\in \mathbb{R}^{H \times W \times C}$, we firstly obtain the k-space spectrum via FFT, i.e., $K_{in}=FFT(S_{in})$. Then, we send $K_{in}$ and $S_{in}$ to k-space branch and image branch, obtaining $K_{out}$ and $S_{out}$ as follows: 
\begin{equation}
\begin{split}
K_{out}&=HiRe \left [ k\-/scan\& Hi\-/scan(K_{in}) \right ]  \\
S_{out}&=HiRe \left [ s\-/scan\& Hi\-/scan(S_{in}) \right ],
\label{eq14}
\end{split}
\end{equation}
where $k\-/scan (\cdot)$ is our customized circular scanning for k-space spectrum unfolding. $s\-/scan (\cdot)$ is the vanilla Mamba scanning applied for the image space. $Hi\-/scan (\cdot)$ is our designed hierarchical scanning strategy to improve the scanning efficiency of both $k\-/scan (\cdot)$ and $s\-/scan (\cdot)$. The symbol $\&$ denotes the combined effect of two operations. $HiRe (\cdot)$ is the process of hierarchical representation. 

Finally, we employ IFFT to $K_{out}$ and aggregate the resulted features with $S_{out}$ via element-wise summation, obtaining the output features $F_{out}$ as follows:
\begin{equation}
\begin{split}
F_{out}=S_{out}+IFFT(K_{out}).
\label{eq15}
\end{split}
\end{equation}
Next, we will provide a detailed introduction to the designed k-space circular scanning ($k\-/scan (\cdot)$), the proposed hierarchical scanning for dual-domains ($Hi\-/scan (\cdot)$), and the detailed pipeline of the hierarchical representation process ($HiRe (\cdot)$) in the following sections.
\subsubsection{Circular Scanning for K-space Domain ($k\-/scan (\cdot)$)}
Vanilla Mamba adopts four-directions scanning paths, i.e., $\delta_{j\in \left \{1,2,3,4 \right\}}$, to unfold images row by row or column by column. However, the k-space spectrum has a concentric circular structure, with low frequencies near the center and higher frequencies around the periphery, as shown in Fig. \ref{Fig.2} (c). Directly applying the standard Mamba scanning disrupts this unique arrangement, complicating k-space global modeling. To address this, we propose a circular scanning strategy. This method unfolds the k-space spectrum along four circular paths, i.e., $\xi_{i\in \left \{1,2,3,4 \right\}}$, obtaining four 1D sequences from low to high frequencies. In this way, we can not only model the association between high and low frequencies but also take into account the periodicity of the k-space.
\subsubsection{Hierarchical Scanning for Dual Domains ($Hi\-/scan (\cdot)$)} As outlined in ~\cite{zhang2024hrvmamba,yuan2024remamba}, the multi-path high-resolution scanning of Mamba results in long-range forgetting and high computational overhead. To tackle this issue, we propose a simple yet effective hierarchical scanning strategy for both k-space and image space unfolding. We divide the four scanning directions into two groups: one serves as the dominant path to unfold at high resolution (HR) and the other three obtain 1D sequences using the downsampled low-resolution (LR) feature maps. In this way, we not only preserve the pixel-level fine-grained features, but also shorten the sequence length to alleviate the long-range forgetting with reduced computational costs.
\begin{figure*}[!t]
\centerline{\includegraphics[width=1.0\linewidth]{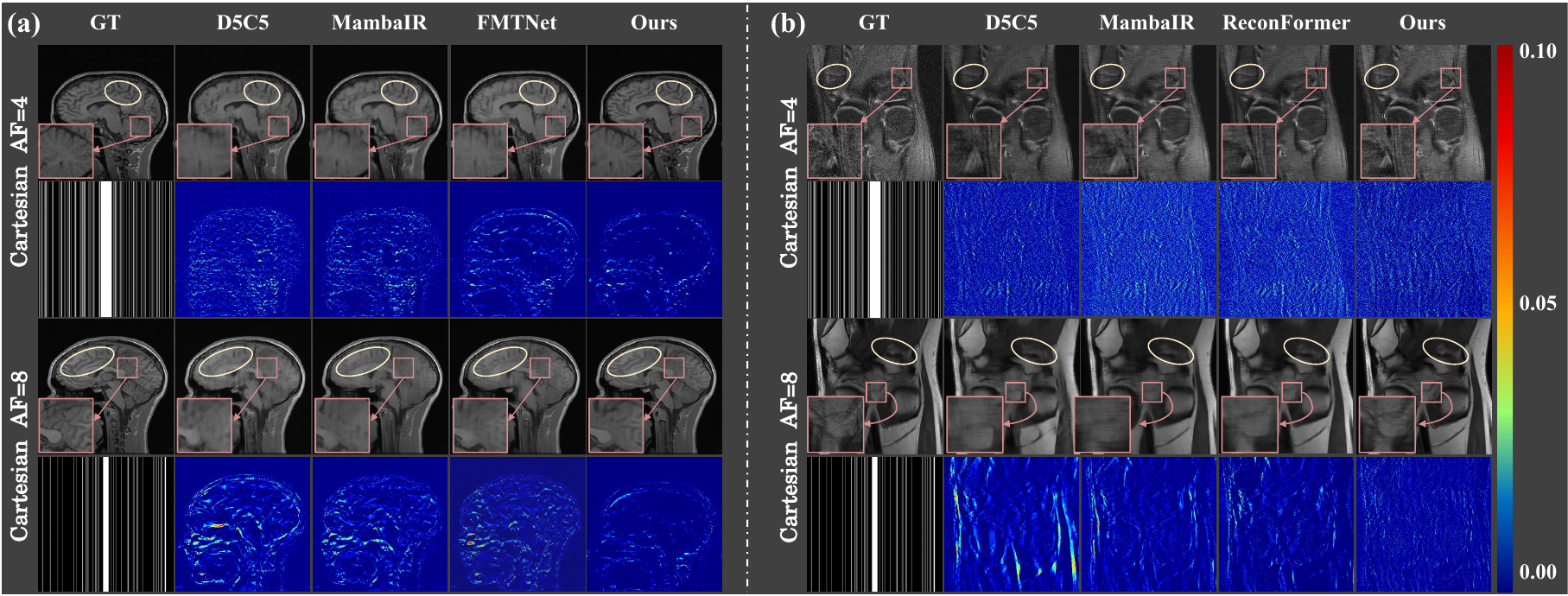}}
\vspace{-0.5 em}
\caption{Visualization comparison on the single-coil datasets, including (a) the CC359 dataset and (b) the fastMRI dataset. The first row of each subplot shows the magnified results of the corresponding red boxes, while the second row shows the error maps. The yellow ellipses highlight the details in the results.}
\label{Fig.4}
\vspace{-0.8 em}
\end{figure*}
\begin{table*}[t]
\renewcommand{\arraystretch}{0.9}
    \centering
    \small
    \setlength{\tabcolsep}{1.9 mm}
    \vspace{-0.5 em}
    \caption{Performance comparison of MRI reconstruction under $4 \times$ and $8 \times$ Acceleration Factor (AF) on the CC359 validation dataset. \\$\mathcal{C}$: CNN-based methods. $\mathcal{T}$: transformer-based methods. $\mathcal{M}$: Mamba-based methods.}
\begin{tabular}{l|c|cc|cc|cc}
\toprule
\rowcolor[HTML]{FFFFFF} 
\cellcolor[HTML]{FFFFFF}                         & \cellcolor[HTML]{FFFFFF}                       & \multicolumn{2}{c|}{\cellcolor[HTML]{FFFFFF}NMSE $\downarrow$} & \multicolumn{2}{c|}{\cellcolor[HTML]{FFFFFF}SSIM $\uparrow$} & \multicolumn{2}{c}{\cellcolor[HTML]{FFFFFF}PSNR $\uparrow$} \\ \cmidrule{3-8} 
\rowcolor[HTML]{FFFFFF} 
\multirow{-2}{*}{\cellcolor[HTML]{FFFFFF}Method} & \multirow{-2}{*}{\cellcolor[HTML]{FFFFFF}Type} & AF=4                    & AF=8                    & AF=4                    & AF=8                    & AF=4                    & AF=8                   \\ \midrule
\rowcolor[HTML]{FFFFFF} 
CS~\cite{r39}                                               &                                                & 0.0481 ± 0.0016           & 0.1068 ± 0.0049           & 0.7611  ±  0.0065           & 0.6420 ± 0.0117           & 26.50 ± 0.51              & 22.68 ± 0.49             \\ \midrule
\rowcolor[HTML]{FFFFFF} 
UNet-32~\cite{r36}                                          & \cellcolor[HTML]{FFFFFF}                       & 0.0197 ± 0.0019           & 0.0385 ± 0.0058           & 0.8898 ± 0.0079           & 0.8348 ± 0.0159           & 31.54 ± 0.42              & 28.66 ± 0.55             \\
\rowcolor[HTML]{FFFFFF} 
KIKI-Net~\cite{r18}                                         & \cellcolor[HTML]{FFFFFF}                       & 0.0221 ± 0.0044           & 0.0417 ± 0.0106           & 0.8415 ± 0.0251           & 0.7773 ± 0.0328           & 28.97 ± 1.40              & 26.24 ± 1.40             \\
\rowcolor[HTML]{FFFFFF} 
D5C5~\cite{r17D5C5}                                             & \cellcolor[HTML]{FFFFFF}                       & 0.0177 ± 0.0019           & 0.0428 ± 0.0063           & 0.8977 ± 0.0088           & 0.8267 ± 0.0176           & 31.59 ± 0.46              & 28.20 ± 0.53             \\
\rowcolor[HTML]{FFFFFF} 
DCRCN~\cite{r19DCRCN}                                            & \multirow{-4}{*}{\cellcolor[HTML]{FFFFFF}$\mathcal{C}$}    & 0.0119 ± 0.0031           & 0.0291 ± 0.0042           & 0.9100 ± 0.0093           & 0.8649 ± 0.0132           & 32.01 ± 1.52              & 29.49 ± 1.35             \\ \midrule
\rowcolor[HTML]{FFFFFF} 
ViT-Base~\cite{r6}                                         & \cellcolor[HTML]{FFFFFF}                       & 0.0207 ± 0.0020           & 0.0446 ± 0.0075           & 0.8903 ± 0.0093           & 0.8254 ± 0.0167           & 31.33 ± 0.47              & 28.03 ± 0.70             \\
\rowcolor[HTML]{FFFFFF} 
SwinMR~\cite{r25}                                           & \cellcolor[HTML]{FFFFFF}                       & 0.0109 ± 0.0014           & 0.0260 ± 0.0064           & 0.9298 ± 0.0074           & 0.8695 ± 0.0148           & 34.14 ± 0.49              & 30.36 ± 0.60             \\
\rowcolor[HTML]{FFFFFF} 
ReconFormer~\cite{r27}                                      & \cellcolor[HTML]{FFFFFF}                       & 0.0108 ± 0.0014           & 0.0276 ± 0.0044           & 0.9297 ± 0.0077           & 0.8650 ± 0.0152           & 34.16 ± 0.51              & 30.11 ± 0.57             \\
\rowcolor[HTML]{FFFFFF} 
FMT-Net~\cite{r26}                                          & \multirow{-4}{*}{\cellcolor[HTML]{FFFFFF}$\mathcal{T}$}    & 0.0058 ± 0.0015           & 0.0166 ± 0.0046           & 0.9366 ± 0.0067           & 0.8665 ± 0.1611           & 36.23 ± 0.65              & 31.32 ± 0.48             \\ \midrule
\rowcolor[HTML]{FFFFFF} 
MambaIR~\cite{r31}                                          & \cellcolor[HTML]{FFFFFF}                       & 0.0087 ± 0.0012           & 0.0207 ± 0.0038           & 0.9506 ± 0.0066           & 0.9005 ± 0.0138           & 35.13 ± 0.54              & 31.37 ± 0.64             \\
\rowcolor[HTML]{FFFFFF} 
VMambaIR~\cite{r32}                                         & \multirow{-2}{*}{\cellcolor[HTML]{FFFFFF}$\mathcal{M}$}    & 0.0119 ± 0.0018           & 0.0270 ± 0.0048           & 0.9352 ± 0.0079           & 0.8807 ± 0.0149           & 33.74 ± 0.60              & 30.22 ± 0.65             \\
\rowcolor[HTML]{EFEFEF} 
\textbf{Ours}                                    & \textbf{}                                      & \textbf{0.0034 ± 0.0012}  & \textbf{0.0130 ± 0.0048}  & \textbf{0.9729 ± 0.0065}  & \textbf{0.9295 ± 0.0176}  & \textbf{39.32 ± 1.24}     & \textbf{33.61 ± 1.41}    \\ \bottomrule
\end{tabular}
    \label{table1}
\vspace{-1.5 em}
\end{table*}
Specifically, given the input feature $K_{in}, S_{in}\in \mathbb{R}^{H \times W \times C}$, we firstly use depthwise convolution for downsampling ($down(\cdot)$). Hence, we obtain hierarchical feature maps as follows:
\begin{equation}
\begin{split}
K_{HR}&=K_{in}, K_{LR}=down_{s}(K_{in}) \\
S_{HR}&=S_{in}, S_{LR}=down_{s}(S_{in}),
\label{eq16}
\end{split}
\end{equation}
where $s$ denotes the convolution stride to control the resolution of $K_{LR}$ and $S_{LR}$. Subsequently, we unfold feature maps to obtain 1D sequences $K_{i \in \left \{ 1,2,3,4 \right \}}$ and $S_{j \in \left \{ 1,2,3,4 \right \}}$ as:
\begin{equation}
\begin{split}
K_{1}&=\xi_{1}(K_{HR}), K_{i}=\xi_{i}(K_{LR}),i\in \left \{2,3,4 \right\} \\
S_{1}&=\delta_{1}(S_{HR}), S_{j}=\delta_{j}(S_{LR}),j\in \left \{2,3,4 \right\}, 
\label{eq17}
\end{split}
\end{equation}
where $\xi_{i\in \left \{1,2,3,4 \right\}}$ are the designed k-space circular scanning paths, and $\delta_{j\in \left \{1,2,3,4 \right\}}$ are the vanilla Mamba scanning paths.
\subsubsection{Hierarchical Representation ($HiRe (\cdot)$)}
To further capture long-range dependencies and extract both fine-grained and coarse global features, we propose a hierarchical representation to process the obtained 1D sequences of varying lengths. Specifically, for $K_{i\in \left \{1,2,3,4 \right\}}$ and $S_{j\in \left \{1,2,3,4 \right\}}$, we use different S6 blocks to process the high-resolution and low-resolution sequences separately. This process can be written as follows:
\begin{equation}
\begin{split}
K_{1}^{'}&=S6_{HR}(K_{1}), K_{i}^{'}=S6_{LR}(K_{i}),i\in \left \{2,3,4 \right\} \\
S_{1}^{'}&=S6_{HR}(S_{1}), S_{j}^{'}=S6_{LR}(S_{j}),j\in \left \{2,3,4 \right\},
\label{eq18}
\end{split}
\end{equation}
where $S6_{HR}(\cdot)$ and $S6_{LR}(\cdot)$ are the operation of S6, which process the 1D sequences as illustrated in Eq.(\ref{eq7}). Subsequently, we convert back the obtained sequences $K^{'}_{i\in \left \{1,2,3,4 \right\}}$ and $S^{'}_{j\in \left \{1,2,3,4 \right\}}$ into 2D feature maps as follows:
\begin{equation}
\begin{split}
K_{i}^{*}&=\xi_{i}^{'}(K_{i}^{'}), i\in \left \{1,2,3,4 \right\} \\
S_{j}^{*}&=\delta_{j}^{'}(S_{j}^{'}), j\in \left \{1,2,3,4 \right\},
\label{eq19}
\end{split}
\end{equation}
where $\xi_{i}^{'}$ and $\delta_{j}^{'}$ are the inverse transformation of $\xi_{i}$ and $\delta_{j}$, respectively. Finally, we interpolate the downsampled maps, i.e., $K_{i \in[2,3,4]}^{*}$ and $S_{j \in[2,3,4]}^{*}$, obtaining $K_{out}$ and $S_{out}$:
\begin{equation}
\begin{split}
K_{out}&=K_{1}^{*}+up( {\textstyle \sum_{i=2}^{i=4}K_{i}^{*}}) \\
S_{out}&=S_{1}^{*}+up( {\textstyle \sum_{j=2}^{j=4}S_{j}^{*}}),
\label{eq20}
\end{split}
\end{equation}
where $up(\cdot)$ denotes the up-sampling interpolation.

\subsection{Local Enhancement Module (LEM)}
Current Mamba methods aggregate features via linear scaling and employ Multilayer Perceptron (MLP) for feature propagation. They overlook spatially-varying representations and result in limited local diversity, hindering accurate MRI. To address this, we propose a Local Enhancement Module (LEM) for selective contributions of local features. 

As shown in Fig.\ref{Fig.3}, LEM treats features from local convolutions as coordinates, multiplying them by a pixel-wise gating mask. Specifically, given input feature $f_{l}\in \mathbb{R}^{H \times W \times C}$, we first expand it to a channel dimension of $2C$ through a $1 \times 1$ convolution. Then we split it into $f_{1}\in \mathbb{R}^{H \times W \times C}$ and $f_{2}\in \mathbb{R}^{H \times W \times C}$. Subsequently, $f_{1}$ and $f_{2}$ are processed in separate branches. In the first branch, $f_{1}$ undergoes a $1 \times 1$ convolution, a $3 \times 3$ depth-wise convolution, and a channel attention block (CAB) to generate the coordinates $C_{r}$:
\begin{equation}
C_{r}=CAB(dwconv(conv(f_{1}))).
\label{eq21}
\end{equation}

In the second branch, $f_{2}$ is processed via a pixel-wise linear projection, somewhat similar to the MLP layer. Unlike a standard MLP, here we apply GELU activation only at the end of the linear projection to generate the gate weights $W_{g}$:
\vspace{-0.5em}
\begin{equation}
W_{g}=GELU(LN(f_{2})).
\label{eq22}
\end{equation}

Finally, the output features $f^{'}_{l}$ can be obtained as follows:
\vspace{-0.5em}
\begin{equation}
f_{l}^{'}=conv(W_{g}\odot C_{r}),
\label{eq23}
\end{equation}
where $\odot$ represents the Hadamard product operation.

\section{EXPERIMENTS}
\subsection{Experimental Settings}
\subsubsection{Datasets and Evaluation Metrics}
We conducted extensive experiments to evaluate our DH-Mamba on three public datasets, including CC359 \cite{r35}, fastMRI \cite{r36}, and SKM-TEA \cite{r37}. The CC359 dataset is acquired from clinical MR scanners, providing 35 complex-valued single-coil brain MRI scans. Each subject contains approximately 180 cross-sectional images with the matrix of size $256\times256$. Following the official split of CC359, we randomly selected a training set comprising 4,524 slices from 25 subjects, and a test set consisting of 1,700 slices from an additional 10 subjects; The fastMRI dataset is the largest open-access raw MRI dataset, providing 1,172 complex-valued single-coil coronal proton density (PD)-weighted knee MRI scans. Each scan provides approximately 35 coronal cross-sectional knee images with the matrix of size $320\times320$. Following \cite{r36}, we partition this dataset into 973 scans for training and 199 scans (fastMRI validation dataset) for testing; The SKM-TEA raw data provides 155 complex-valued multi-coil T2-weighted knee MRI scans, and each subject provides approximately 160 cross-sectional knee images with the matrix of size $512\times512$. Following the well-established ReconFormer ~\cite{r27}, we used 124, 10, and 21 volumes for training, validation, and testing, respectively.

Normalized mean square error (NMSE), structural index similarity (SSIM), and peak signal-to-noise ratio (PSNR) are used as evaluation metrics for comparison.
\begin{table*}[t]
\renewcommand{\arraystretch}{0.9}
    \centering
    \small
    \setlength{\tabcolsep}{1.9 mm}
    \caption{Performance comparison of MRI reconstruction under $4 \times$ and $8 \times$ Acceleration Factor (AF) on the fastMRI validation dataset. \\$\mathcal{C}$: CNN-based methods. $\mathcal{T}$: transformer-based methods. $\mathcal{M}$: Mamba-based methods.}
\begin{tabular}{l|c|cc|cc|cc}
\toprule
\rowcolor[HTML]{FFFFFF} 
\cellcolor[HTML]{FFFFFF}                         & \multicolumn{1}{l|}{\cellcolor[HTML]{FFFFFF}}                       & \multicolumn{2}{c|}{\cellcolor[HTML]{FFFFFF}NMSE $\downarrow$} & \multicolumn{2}{c|}{\cellcolor[HTML]{FFFFFF}SSIM $\uparrow$} & \multicolumn{2}{c}{\cellcolor[HTML]{FFFFFF}PSNR $\uparrow$} \\ \cmidrule{3-8} 
\rowcolor[HTML]{FFFFFF} 
\multirow{-2}{*}{\cellcolor[HTML]{FFFFFF}Method} & \multicolumn{1}{l|}{\multirow{-2}{*}{\cellcolor[HTML]{FFFFFF}Type}} & AF=4                    & AF=8                    & AF=4                    & AF=8                    & AF=4                    & AF=8                   \\ \midrule
\rowcolor[HTML]{FFFFFF} 
CS~\cite{r39}                                               &                                                                     & 0.0583 ± 0.0420           & 0.0903 ± 0.0440           & 0.5736 ± 0.1220           & 0.4870 ± 0.1330           & 29.54 ± 3.29              & 26.99 ± 2.42             \\ \midrule
\rowcolor[HTML]{FFFFFF} 
UNet-32~\cite{r36}                                          & \cellcolor[HTML]{FFFFFF}                                            & 0.0337 ± 0.0247           & 0.0477 ± 0.0280           & 0.7248 ± 0.1312           & 0.6570 ± 0.1482           & 31.99 ± 3.34              & 30.02 ± 2.67             \\
\rowcolor[HTML]{FFFFFF} 
KIKI-Net~\cite{r18}                                         & \cellcolor[HTML]{FFFFFF}                                            & 0.0353 ± 0.0270           & 0.0546 ± 0.0280           & 0.7172 ± 0.1340           & 0.6355 ± 0.1460           & 31.87 ± 3.47              & 29.27 ± 2.44             \\
\rowcolor[HTML]{FFFFFF} 
D5C5~\cite{r17D5C5}                                             & \cellcolor[HTML]{FFFFFF}                                            & 0.0332 ± 0.0260           & 0.0512 ± 0.0280           & 0.7256 ± 0.1360           & 0.6457 ± 0.1470           & 32.25 ± 3.65              & 29.65 ± 2.63             \\
\rowcolor[HTML]{FFFFFF} 
DCRCN~\cite{r19DCRCN}                                            & \multirow{-4}{*}{\cellcolor[HTML]{FFFFFF}$\mathcal{C}$}                         & 0.0351 ± 0.0150           & 0.0443 ± 0.0138           & 0.7332 ± 0.0535           & 0.6635 ± 0.0613           & 32.18 ± 1.65              & 30.76 ± 1.41             \\ \midrule
\rowcolor[HTML]{FFFFFF} 
ViT-Base~\cite{r6}                                         & \cellcolor[HTML]{FFFFFF}                                            & 0.0342 ± 0.0267           & 0.0460 ± 0.0288           & 0.7206 ± 0.1362           & 0.6578 ± 0.1493           & 32.10 ± 3.59              & 30.28 ± 2.83             \\
\rowcolor[HTML]{FFFFFF} 
SwinMR~\cite{r25}                                           & \cellcolor[HTML]{FFFFFF}                                            & 0.0342 ± 0.0270           & 0.0476 ± 0.031            & 0.7213 ± 0.137            & 0.6537 ± 0.1510           & 32.14 ± 3.64              & 30.21 ± 2.95             \\
\rowcolor[HTML]{FFFFFF} 
ReconFormer~\cite{r27}                                      & \cellcolor[HTML]{FFFFFF}                                            & 0.0320 ± 0.0254           & 0.0431 ± 0.0293           & 0.7365 ± 0.1364           & 0.6672 ± 0.1528           & 32.62 ± 3.79              & 30.82 ± 3.14             \\
\rowcolor[HTML]{FFFFFF} 
FMT-Net~\cite{r26}                                          & \multirow{-4}{*}{\cellcolor[HTML]{FFFFFF}$\mathcal{T}$}                         & 0.0441 ± 0.0199           & 0.0535 ± 0.0223           & 0.7364 ± 0.1561           & 0.6661 ± 0.1661           & 32.56 ± 2.98              & 31.06 ± 2.88             \\ \midrule
\rowcolor[HTML]{FFFFFF} 
MambaIR~\cite{r31}                                          & \cellcolor[HTML]{FFFFFF}                                            & 0.0317 ± 0.0262           & 0.0415 ± 0.0225           & 0.8172 ± 0.1442           & 0.7691 ± 0.1365           & 32.51 ± 2.53              & 30.96 ± 2.86             \\
\rowcolor[HTML]{FFFFFF} 
VMambaIR~\cite{r32}                                         & \multirow{-2}{*}{\cellcolor[HTML]{FFFFFF}$\mathcal{M}$}                         & 0.0453 ± 0.0265           & 0.0544 ± 0.0261           & 0.7856 ± 0.1445           & 0.6882 ± 0.1453           & 32.46 ± 2.44              & 30.54 ± 2.18             \\
\rowcolor[HTML]{EFEFEF} 
\textbf{Ours}                                             &                                                                     & \textbf{0.0308 ± 0.0211}  & \textbf{0.0403 ± 0.0198}  & \textbf{0.8211 ± 0.0156}  & \textbf{0.7741 ± 0.1376}  & \textbf{32.76 ± 2.35}     & \textbf{31.17 ± 2.07}    \\ \bottomrule
\end{tabular}
    \label{table2}
\vspace{-1.0 em}
\end{table*}
\begin{table*}[t]
\renewcommand{\arraystretch}{0.9}
    \centering
    \small
    \setlength{\tabcolsep}{1.9 mm}
    \caption{Performance comparison of MRI reconstruction under $4 \times$ and $8 \times$ Acceleration Factor (AF) on the multi-coil SKM-TEA dataset. \\$\mathcal{C}$: CNN-based methods. $\mathcal{T}$: transformer-based methods. $\mathcal{M}$: Mamba-based methods.}
\begin{tabular}{l|c|cc|cc|cc}
\toprule
                          &                        & \multicolumn{2}{c|}{NMSE $\downarrow$}                           & \multicolumn{2}{c|}{SSIM $\uparrow$}                           & \multicolumn{2}{c}{PSNR $\uparrow$}                      \\ \cmidrule{3-8} 
\multirow{-2}{*}{Method} & \multirow{-2}{*}{Type} & AF=4                     & AF=8                     & AF=4                     & AF=8                     & AF=4                  & AF=8                  \\ \midrule
UNet-32~\cite{r36}                   &                        & 0.0204 ± 0.0032          & 0.0270 ± 0.0041          & 0.8469 ± 0.0262          & 0.7904 ± 0.0030          & 33.91 ± 1.39          & 31.44 ± 1.24          \\
KIKI-Net~\cite{r18}                  &                        & 0.0196 ± 0.0028          & 0.0271 ± 0.0040          & 0.8577 ± 0.0264          & 0.7941 ± 0.0031          & 34.26 ± 1.42          & 31.42 ± 1.28          \\
D5C5~\cite{r17D5C5}                      & \multirow{-3}{*}{$\mathcal{C}$}    & 0.0188 ± 0.0033          & 0.0257 ± 0.0039          & 0.8648 ± 0.0258          & 0.8030 ± 0.0031          & 34.63 ± 1.44          & 31.89 ± 1.30          \\ \midrule
SwinMR~\cite{r25}                    &                        & 0.0192 ± 0.0025          & 0.0256 ± 0.0038          & 0.8597 ± 0.0249          & 0.8022 ± 0.0028          & 34.45 ± 1.37          & 31.94 ± 1.27          \\
Reconformer~\cite{r27}               & \multirow{-2}{*}{$\mathcal{T}$}    & 0.0179 ± 0.0029          & 0.0239 ± 0.0029          & 0.8730 ± 0.0252          & 0.8158 ± 0.0030          & 35.06 ± 1.45          & 32.51 ± 1.32          \\ \midrule
MambaIR~\cite{r31}                   & $\mathcal{M}$                      & 0.0186 ± 0.0028          & 0.0245 ± 0.0040          & 0.8673 ± 0.0248          & 0.8032 ± 0.0027          & 34.88 ± 1.42          & 32.48 ± 1.30          \\
\rowcolor[HTML]{EFEFEF} 
\textbf{Ours}             & \textbf{}              & \textbf{0.0156 ± 0.0026} & \textbf{0.0020 ± 0.0032} & \textbf{0.9154 ± 0.0232} & \textbf{0.8979 ± 0.0025} & \textbf{35.43 ± 1.40} & \textbf{32.97 ± 1.28} \\ \bottomrule
\end{tabular}
    \label{table3}
\vspace{-1.2 em}
\end{table*}
\begin{figure}[!t]
\centerline{\includegraphics[width=1.0\linewidth]{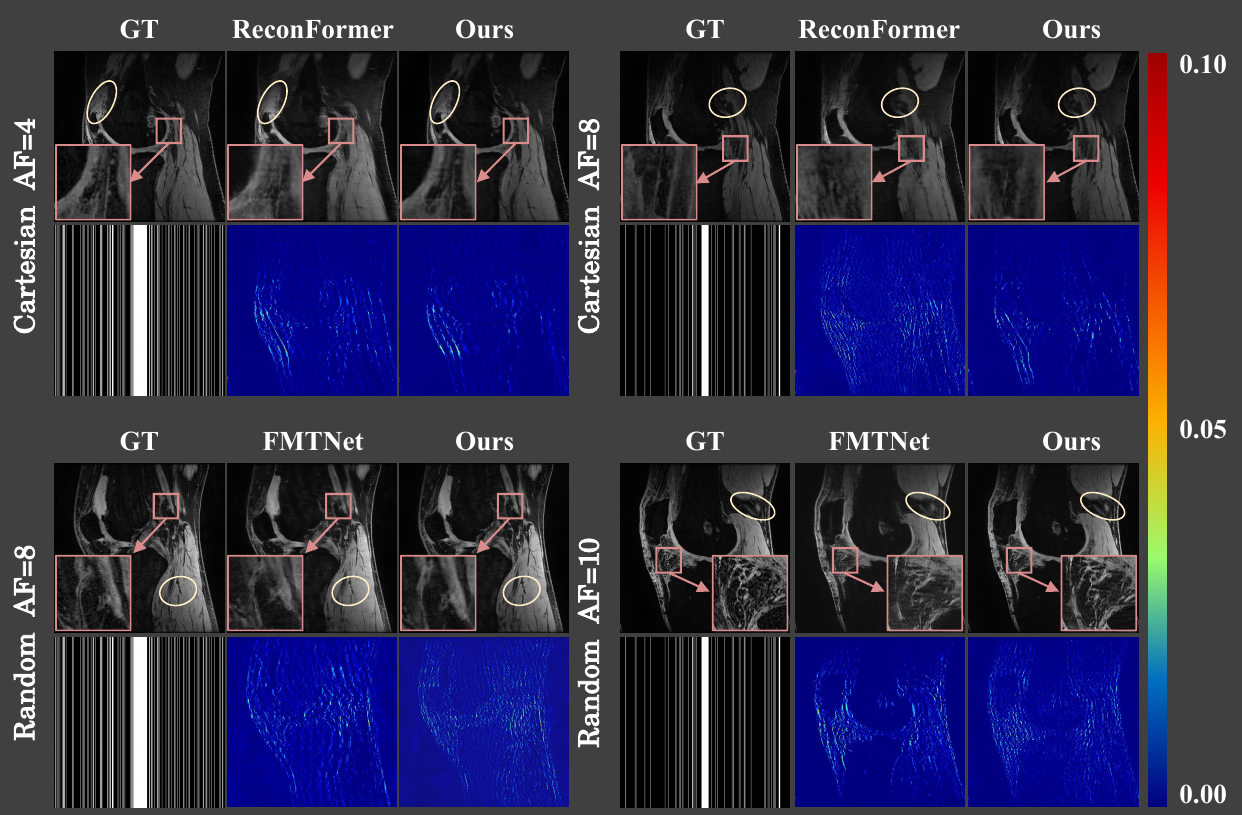}}
\caption{Visualization comparison on the multi-coil SKM-TEA dataset.}
\label{Fig.5}
\vspace{-1.5 em}
\end{figure}
\subsubsection{Mask Generation}
In our main comparisons, the inputs are generated by randomly undersampling the k-space data using 1D Cartesian masks, similar to the approach employed in the fastMRI challenge \cite{r36}. Specifically, when the acceleration factor (AF) equals $4$, the fully-sampled central region includes $8\%$ of all k-space lines; when it equals $8$, $4\%$ of all k-space lines are included. The remaining k-space lines are included uniformly at random, with the probability set so that, on average, the undersampling mask achieves the desired acceleration factor. In our mask experiments, we followed the methodology outlined in \cite{r38}, utilizing radial and random masks to create a wider variety of under-sampling patterns.
\subsubsection{Implementation Details}
In our DH-Mamba, the number of DHM Groups $M$ and the number of DHM Blocks $N$ are set to $(6,6)$. For the DHM module, we set the scale of the hierarchical scanning as $s=2$, and set the state size of the S6 operation as $H=16$. The channel dimension $C$ is empirically set as $64$. During training, we employ the AdamW optimizer, setting $\beta_{1}=0.9, \beta_{2}=0.99$, and momentum $=0$. The initial learning rate is fixed at $1\times 10^{-4}$ and reduces to $1\times 10^{-6} $ using a cosine annealing schedule with a weight decay of $0.05$.
\begin{table*}[h]
\renewcommand{\arraystretch}{0.9}
    \centering
    \small
    \setlength{\tabcolsep}{1.7mm}
    \caption{performance comparison on cc359 using more mask patterns and acceleration factors (af). $\mathcal{I}$: radial mask. $\mathcal{R}$: random mask.}
\begin{tabular}{l|c|cclc|cclc|cclc}
\toprule
                         &                        & \multicolumn{4}{c|}{NMSE $\downarrow$}                                  & \multicolumn{4}{c|}{SSIM $\uparrow$}                                                      & \multicolumn{4}{c}{PSNR $\uparrow$}                                                    \\ \cmidrule{3-14} 
\multirow{-2}{*}{Method} & \multirow{-2}{*}{Mask} & AF=4            & AF=5            & AF=8 & AF=10           & AF=4            & AF=5            & \multicolumn{1}{c}{AF=8} & AF=10           & AF=4           & AF=5           & \multicolumn{1}{c}{AF=8} & AF=10          \\ \midrule
ReconFormer~\cite{r27}              &                        & 0.0024          & 0.0030          &0.0052      & 0.0070          & 0.9754          & 0.9650          &0.9523                          & 0.9230          & 42.52          & 40.18          &38.42                          & 36.15          \\
FMTNet~\cite{r26}                   &                        & 0.0015          & 0.0024          &0.0050      & 0.0068          & 0.9825          & 0.9769          &0.9583                          & 0.9547          & 42.63          & 40.62          & 38.79                         & 36.19          \\
MambaIR~\cite{r31}                  &                        & 0.0037          & 0.0051          &0.0082      & 0.0124          & 0.9691          & 0.9623          & 0.9397                         & 0.9305          & 38.77          & 37.44          & 35.66                         & 33.56          \\
VMambaIR~\cite{r32}                 & \multirow{-4}{*}{$\mathcal{I}$}    & 0.0045          & 0.0059          &0.0088      & 0.0145          & 0.9656          & 0.9583          &0.9288                          & 0.9222          & 37.98          & 36.75          &34.32                          & 32.88          \\
\rowcolor[HTML]{EFEFEF} 
\textbf{Ours}            & \textbf{}              & \textbf{0.0014} & \textbf{0.0020} & \textbf{0.0028}      & \textbf{0.0055} & \textbf{0.9838} & \textbf{0.9794} & \textbf{0.9727}                          & \textbf{0.9615} & \textbf{43.18} & \textbf{41.49} & \textbf{39.96}                          & \textbf{37.13} \\ \midrule
ReconFormer~\cite{r27}             &                        & 0.0022          & 0.0042          &0.0051      & 0.0064          & 0.9612          & 0.9597          &0.9466                          & 0.9347          & 41.45          & 40.31          &39.65                          & 37.14          \\
FMTNet~\cite{r26}                   &                        & 0.0018          & 0.0026          &0.0045      & 0.0053          & 0.9798          & 0.9741          &0.9611                          & 0.9599          & 42.08          & 40.34          &39.86                          & 37.24          \\
MambaIR~\cite{r31}                  &                        & 0.0087          & 0.0102          &0.0142      & 0.0151          & 0.9441          & 0.9371          & 0.9316                         & 0.9170          & 35.09          & 34.39          &33.16                          & 32.70          \\
VMambaIR~\cite{r32}                 & \multirow{-4}{*}{$\mathcal{R}$}    & 0.0095          & 0.0111          &0.0156      & 0.0168          & 0.9403          & 0.9332          &0.9298                          & 0.9106          & 34.67          & 34.02          &32.81                          & 32.22          \\
\rowcolor[HTML]{EFEFEF} 
\textbf{Ours}            & \textbf{}              & \textbf{0.0015} & \textbf{0.0021} & \textbf{0.0022}     & \textbf{0.0045} & \textbf{0.9819} & \textbf{0.9777} & \textbf{0.9765}                         & \textbf{0.9643} & \textbf{42.79} & \textbf{41.41} & \textbf{40.98}      & \textbf{37.94} \\ \bottomrule
\end{tabular}
    \label{table4}
\vspace{-1.5 em}
\end{table*}
\begin{figure}[!t]
\centerline{\includegraphics[width=\columnwidth]{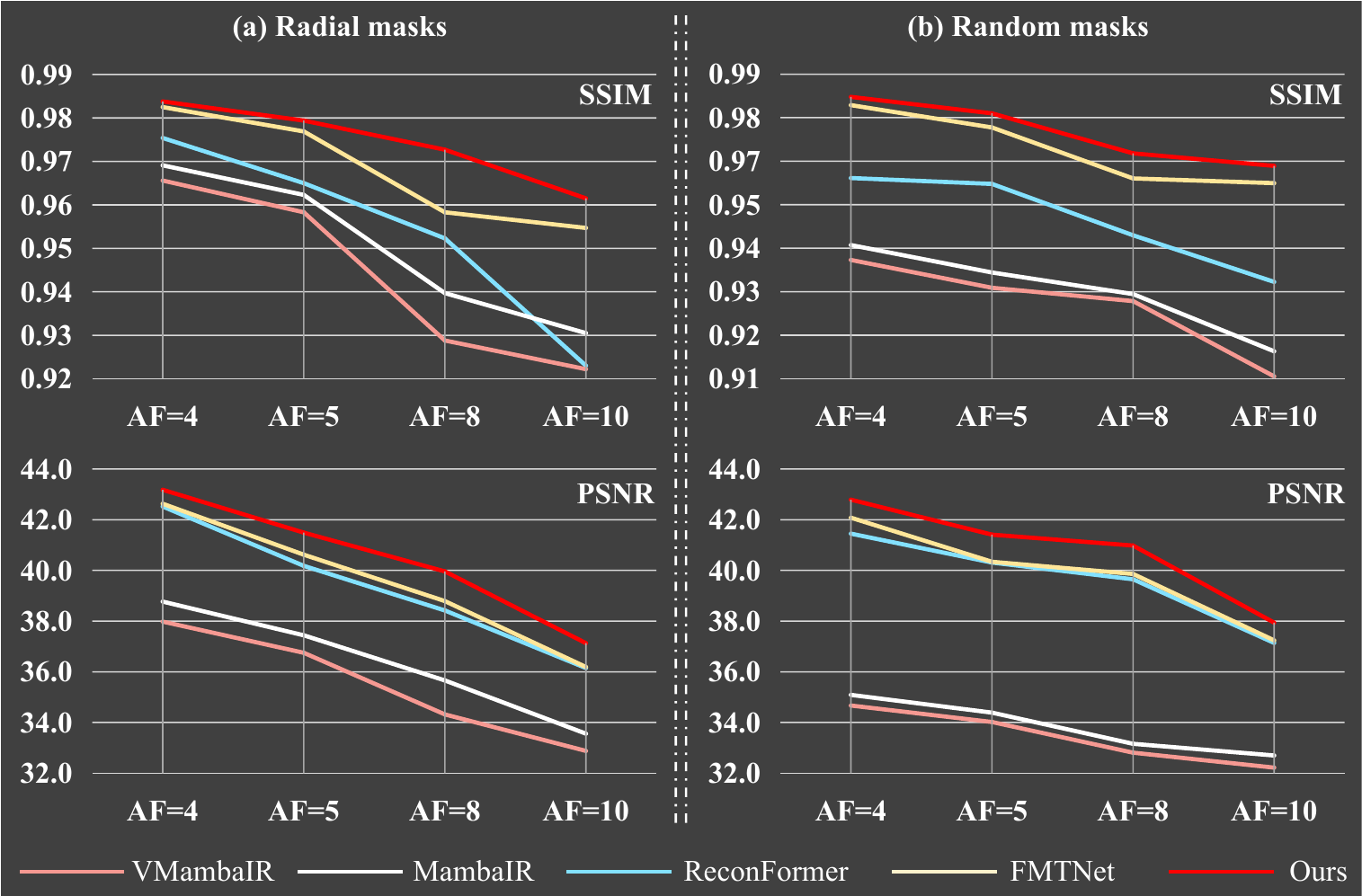}}
\caption{Results of the mask experiments on the CC359. The X-axis is the acceleration factors, while the Y-axis corresponds to various metrics.}
\label{Fig.6}
\vspace{-1.5 em}
\end{figure}
\subsubsection{Comparison Methods}
To verify the effectiveness of our DH-Mamba, we evaluated it against closely related accelerated MRI algorithms, including CNN-based and transformer-based methods. Specifically, we draw comparison with Compressed Sensing (CS)~\cite{r39}, UNet-32~\cite{r36}, KIKI-Net~\cite{r18}, D5C5~\cite{r17D5C5}, DCRCN~\cite{r19DCRCN}, ViT-Base~\cite{r6}, SwinMR~\cite{r25}, Reconformer~\cite{r27}, and FMT-Net~\cite{r26}. Furthermore, to validate the applicability of DH-Mamba for MRI reconstruction, we compared it with two powerful Mamba-based restoration methods, i.e. MambaIR~\cite{r31} and VMambaIR~\cite{r32} and equipped them with a data consistency (DC) layer for fair comparison. Notably, we only draw comparison with the closely related methods that have open-source code, reproducing them using their default parameter settings for a fair comparison.
\subsection{Comparison with State-of-the-arts}
\subsubsection{Single-coil Datasets}
Table \ref{table1} and Table \ref{table2} shows the comparison results of different methods under various acceleration factors on the single-coil CC359 and fastMRI dataset, respectively. As indicated, our proposed DH-Mamba is significantly superior to the CNN methods and shows obvious performance improvements compared to the Transformer and Mamba based state-of-the-arts (SOTAs). Taking the CC359 dataset as an example, our DH-Mamba surpasses DCRCN and FMTNet by $7.31$ dB and $3.09$ dB in terms of PSNR under $4 \times$ AF, respectively. As for the fastMRI dataset, our method shows the superiority of $0.14$ dB and $0.35$ dB over the leading method ReconFormer at $4 \times$ and $8 \times$ AF, respectively. Furthermore, we can find that while MambaIR and VMambaIR serve as powerful backbones for natural image restoration, they achieve poor performance on MRI reconstruction. Taking CC359 dataset as an example, MambaIR and VMambaIR achieves $35.13$ and $33.74$ in terms of PSNR under $4 \times$ AF, respectively. In contrast, the PSNR of DH-Mamba are improved to $39.32$, showing the superiority of $4.19$ dB and $5.58$ dB over these two counterparts respectively. This highlights that our customized methods effectively address the challenges of applying Mamba to MRI reconstruction, resulting in improved performance.
\begin{figure}[!t]
\centerline{\includegraphics[width=1.0\linewidth]{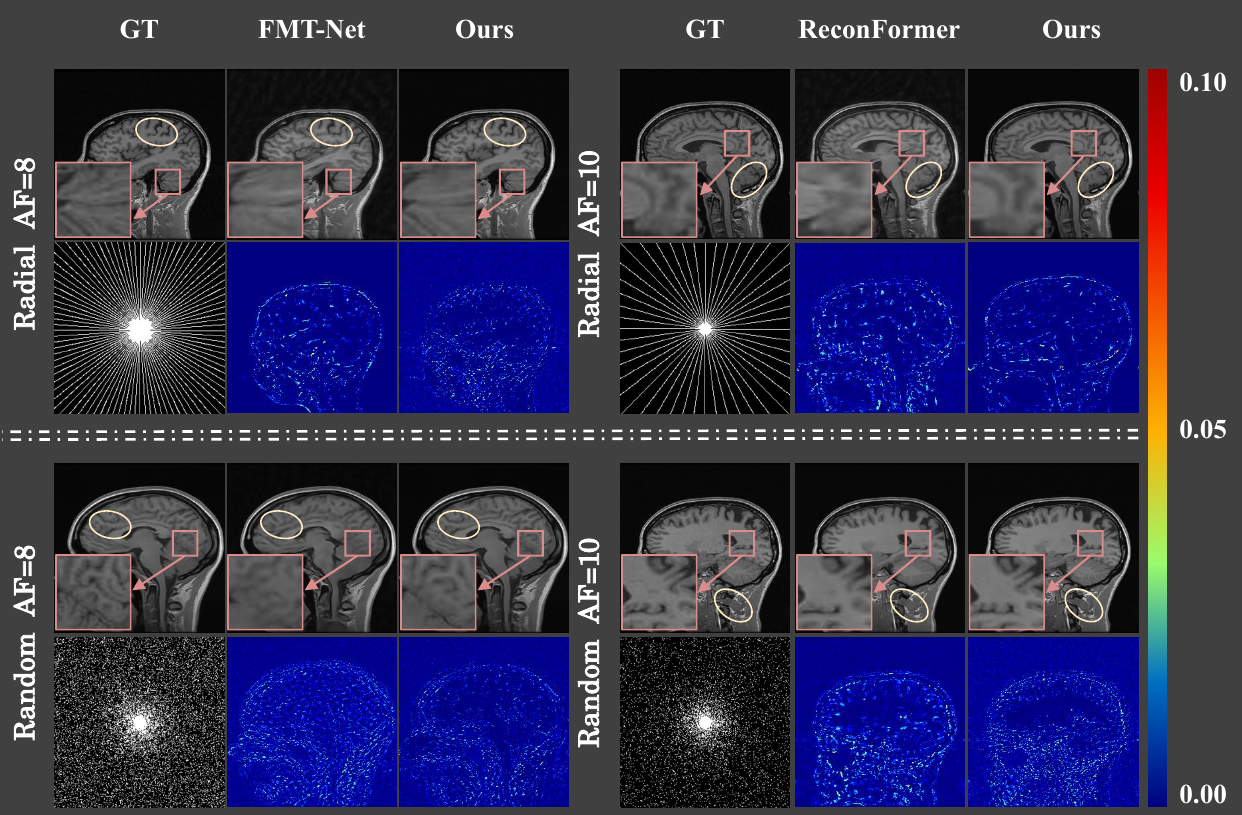}}
\caption{Visualization comparison of different methods under more undersampling patterns. The yellow ellipses and red boxes highlight the details.}
\label{Fig.7}
\vspace{-1.5 em}
\end{figure}

Fig. \ref{Fig.4} presents a qualitative comparison under the Cartesian sampling pattern on CC359 and fastMRI datasets, respectively. In each subplot, the first row magnifies the details within the red boxes, while the second row displays error maps that compare the reconstruction results of different methods with the fully sampled ground truth (GT). As indicated, CNN-based method, i.e., D5C5, struggles to recover the complex texture, resulting in over-smoothing in reconstructed images. Transformer-based methods, including FMTNet and ReconFormer, alleviate this problem, but it is difficult to maintain satisfactory reconstruction results in difficult tasks (AF $=8$). In contrast, our method clearly reconstructs rich details and sharp structures across all datasets and acceleration factors.
\subsubsection{Multi-coil Datasets}
Table \ref{table3} shows the quantitative results on the multi-coil dataset SKM-TEA for AF $=4$ and AF $=8$. It can be seen that our method still maintains high-quality reconstruction on the multi-coil dataset and consistently surpasses other approaches across different acceleration rates. Specifically, our DH-Mamba outperforms Reconformer by $(0.37, 0.0424)$ and $(0.46, 0.0821)$ in terms of (PSNR, SSIM) at $4 \times$ and $8 \times$ acceleration factors, respectively.

Visual comparison results of DH-Mamba and other SOTA algorithms are shown in Fig. \ref{Fig.5}, (please enlarge the images for clear details). Compared to previous algorithms, which suffer from overall blurriness and loss of details, our DH-Mamba achieves significant performance improvement.

\subsubsection{Experiments on Different Masks}
To further validate the effectiveness and robustness of our method, we conducted additional experiments using a wide range of downsampling patterns, as presented in Table \ref{table4} and Fig. \ref{Fig.6}. The results reveal that: (1) Our method consistently outperforms other approaches across various downsampling patterns and acceleration factors, demonstrating its robustness in different scenarios; (2) Even in challenging tasks such as AF $=10$, our method maintains excellent performance. For example, under the random AF $=10$, our method achieves $37.94$ PSNR and $0.9643$ SSIM, exceeding those of ReconFormer by $0.80$ and $0.0296$, respectively; (3) Although MambaIR and VMambaIR have been shown to possess strong capabilities in natural image recovery, they consistently yield suboptimal performance in MRI reconstruction. This highlights the effectiveness of our method, specifically tailored for MRI applications.

In  Fig. \ref{Fig.7}, we present a visual comparison of the output results under radial and random sampling patterns. From the enlarged local comparison, it is evident that our DH-Mamba reconstructs relatively clear structures even in the challenging task of AF $=10$. This reinforces the robustness of our approach in difficult scenarios.
\begin{table}[]
\renewcommand{\arraystretch}{1.1}
    \centering
    \small
    \setlength{\tabcolsep}{0.96mm}
    \vspace{-1.0 em}
    \caption{Ablation studies of different settings of our method. This experiment is conducted on the CC359 dataset under $4 \times$ AF.}
\begin{tabular}{l|cccccccc}
\toprule
\rowcolor[HTML]{FFFFFF} 
Model         & Img.        & Ksp.        & Hi-scan & k-scan   & LEM        & NMSE            & SSIM            & PSNR           \\ \midrule
\rowcolor[HTML]{FFFFFF} 
(a)           &            & \pmb{$\checkmark$}          & \pmb{$\checkmark$}           & \pmb{$\checkmark$}          & \pmb{$\checkmark$}          & 0.0156          & 0.8765          & 36.22          \\
\rowcolor[HTML]{FFFFFF} 
(b)           & \pmb{$\checkmark$}          &            & \pmb{$\checkmark$}           &            & \pmb{$\checkmark$}          & 0.0042          & 0.9652          & 38.34          \\
\rowcolor[HTML]{FFFFFF} 
(c)           & \pmb{$\checkmark$}          & \pmb{$\checkmark$}          &             & \pmb{$\checkmark$}          & \pmb{$\checkmark$}          & 0.0036 & 0.9722 & 39.14          \\
\rowcolor[HTML]{FFFFFF} 
(d)           & \pmb{$\checkmark$}          & \pmb{$\checkmark$}          & \pmb{$\checkmark$}           &            & \pmb{$\checkmark$}          & 0.0040          & 0.9678          & 38.76          \\
\rowcolor[HTML]{FFFFFF} 
(e)           & \pmb{$\checkmark$}          & \pmb{$\checkmark$}          & \pmb{$\checkmark$}           & \pmb{$\checkmark$}          &            & 0.0039          & 0.9706          & 38.82          \\
\rowcolor[HTML]{EFEFEF} 
\textbf{Ours} & \textbf{\pmb{$\checkmark$}} & \textbf{\pmb{$\checkmark$}} & \textbf{\pmb{$\checkmark$}}  & \textbf{\pmb{$\checkmark$}} & \textbf{\pmb{$\checkmark$}} & \textbf{0.0034} & \textbf{0.9729} & \textbf{39.32} \\ \bottomrule
\end{tabular}
    \label{table5}
\vspace{-0.4 em}
\end{table}
\begin{figure}[!t]
\centerline{\includegraphics[width=\columnwidth]{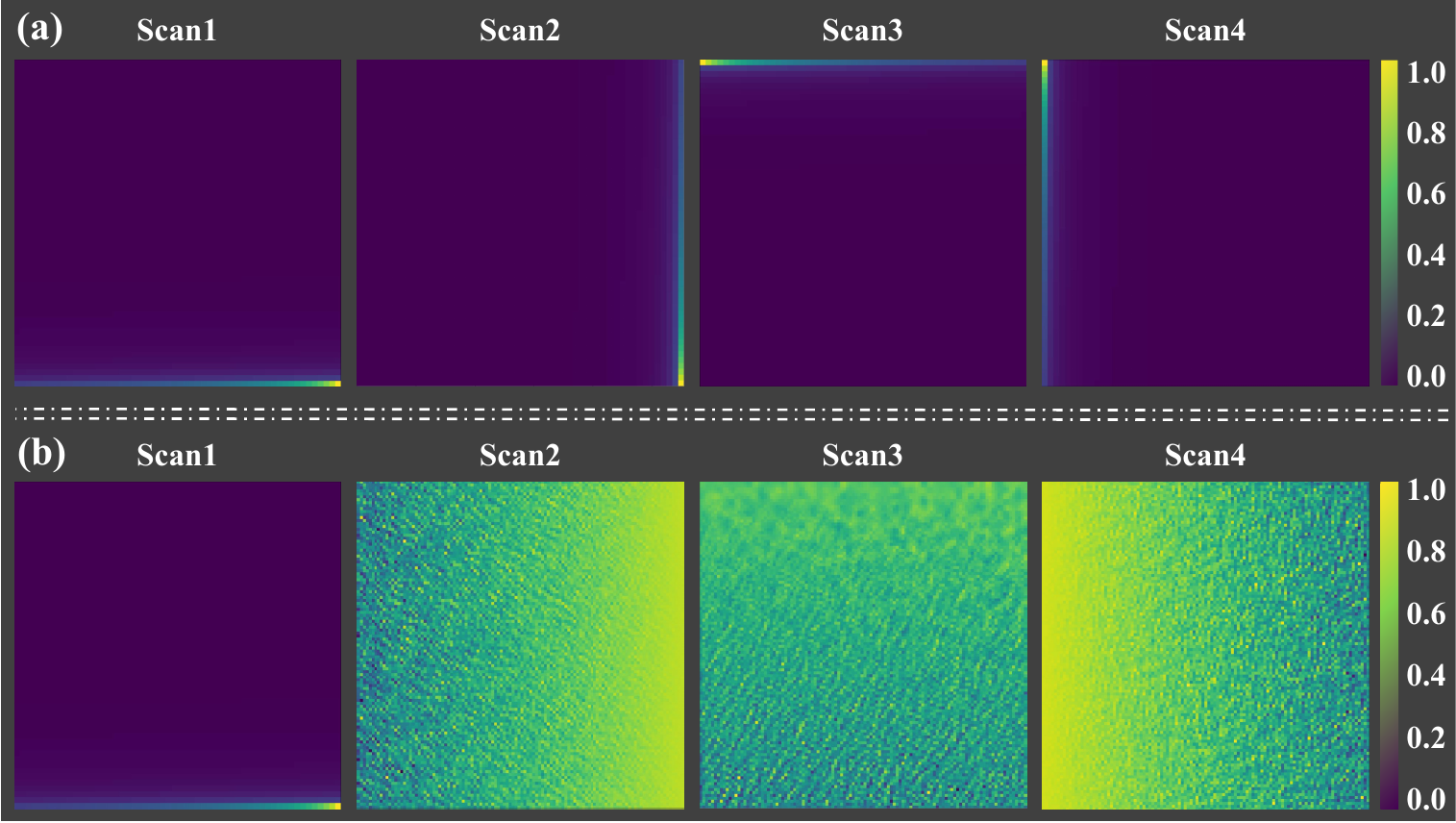}}
\caption{Illustration of the long-range forgetting using general SS2D method (a) and our hierarchical scanning strategy (b).}
\label{Fig.8}
\vspace{-1.0 em}
\end{figure}

\subsection{Ablation Studies}
\subsubsection{Efficacy of Key Components}
We conducted a breakdown ablation experiment on the key modules and strategies of our proposed DH-Mamba to explore their effects and interactions, with the results presented in Table \ref{table5}. The findings indicate that:
(a) Removing the image space branch (Img.) caused a significant drop in PSNR from $39.32$ to $36.22$, highlighting the critical role of image domain recovery in the overall reconstruction process.
(b) Disabling the k-space branch (Ksp.) led to a $0.98$ drop in PSNR and $0.0077$ decrease in SSIM, demonstrating its importance in maintaining frequency domain fidelity.
(c) Replacing our hierarchical scanning (Hi-scan) with the vanilla SS2D approach~\cite{r11} reduced PSNR by $0.18$ dB, indicating that Hi-scan effectively mitigates long-range forgetting for better global modeling.
(d) Swapping our k-space circular scanning (k-scan) with vanilla SS2D resulted in a slight PSNR decrease from $39.32$ to $38.76$, confirming that k-scan helps preserve k-space structure for improved frequency domain modeling.
(e) Removing the Local Enhancement Module (LEM) led to a noticeable performance decline, underscoring its role in enhancing image quality.

\subsubsection{Effectiveness of K-space Circular Scanning}
To further analyze the effectiveness of our proposed k-space circular scanning (k-scan), we compared it with several popular strategies: SS2D~\cite{r11}, Window-SS2D~\cite{r33}, and Continuous-SS2D~\cite{r15}. As shown in Table \ref{table6}, vanilla SS2D achieved only $38.76$ PSNR. Although Window-SS2D and Continuous-SS2D improve SS2D under consideration of image properties, they overlook the unique structure of k-space, resulting in suboptimal results. In contrast, our k-scan achieved $39.32$ PSNR and $0.9729$ SSIM. This highlights that our k-scan is tailored for k-space unfolding, preserving the relationships among different frequencies and facilitating global modeling.
\begin{table}[!t]
\renewcommand{\arraystretch}{0.94}
    \centering
    \small
    \setlength{\tabcolsep}{1.9mm}
    \vspace{-0.90 em}
    \caption{More Detailed ablation analysis of our k-scan and LEM. This experiment is conducted on the CC359 dataset under $4 \times$ AF.}
\begin{tabular}{lc|ccc}
\toprule
\multicolumn{1}{l|}{Model}                       & Module                              & NMSE                           & SSIM                           & PSNR                          \\ \midrule
\multicolumn{1}{l|}{SS2D~\cite{r11}}                        &                                   & 0.0040          & 0.9678          & 38.76                         \\
\multicolumn{1}{l|}{Window-SS2D~\cite{r33}}                 &                                   & \cellcolor[HTML]{FFFFFF}0.0039 & \cellcolor[HTML]{FFFFFF}0.9687 & \cellcolor[HTML]{FFFFFF}38.85 \\
\multicolumn{1}{l|}{Continuous-SS2D~\cite{r15}}              & \multirow{-3}{*}{k-scan}             & \cellcolor[HTML]{FFFFFF}0.0039 & \cellcolor[HTML]{FFFFFF}0.9690 & \cellcolor[HTML]{FFFFFF}38.86 \\ \midrule
\multicolumn{1}{l|}{MLP~\cite{r11}}                         &                                   & 0.0039                         & 0.9706                         & 38.82                         \\
\multicolumn{1}{l|}{CAB~\cite{r31}}                         & \multirow{-2}{*}{LEM}             & 0.0037                         & 0.9716                         & 39.00                         \\ \midrule
\rowcolor[HTML]{EFEFEF} 
\multicolumn{2}{l|}{\cellcolor[HTML]{EFEFEF}\textbf{Our DH-Mamba (k-scan + LEM)}} & \textbf{0.0034}                & \textbf{0.9729}                & \textbf{39.32}                \\ \bottomrule
\end{tabular}
    \label{table6}
\vspace{-0.4 em}
\end{table}
\begin{figure}[!t]
\centerline{\includegraphics[width=\columnwidth]{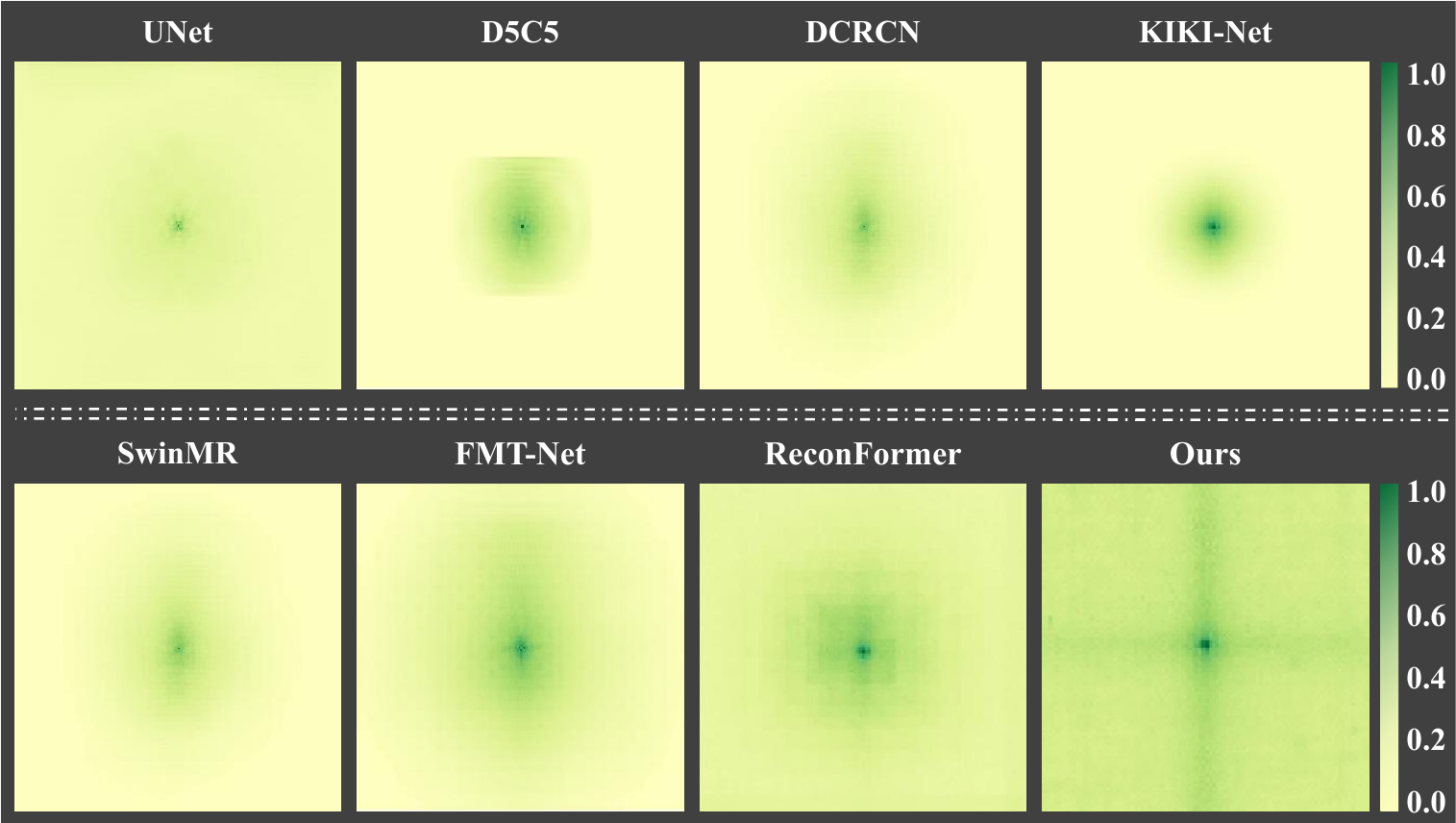}}
\caption{Effective receptive field (ERF) comparisons. A lager ERF is indicated by a more extensively distributed dark area.}
\label{Fig.9}
\vspace{-1.2 em}
\end{figure}
\subsubsection{Effectiveness of Local Enhancement Module}
To evaluate the performance of our proposed Local Enhancement Module (LEM), we assessed it against two widely used feed-forward networks in Mamba-based methods: Multilayer Perceptron (MLP)\cite{r11} and Channel Attention Block (CAB)\cite{r31}. The results in Table \ref{table6} show that both MLP and CAB achieved unsatisfactory results. Specifically, MLP and CAB gained $38.82$ and $39.00$ in terms of PSNR, respectively. The results indicates their limited effectiveness in enhancing Mamba's feature representation. In contrast, our LEM significantly improved local diversity, boosting the reconstruction performance to $39.32$ (PSNR) and $0.9729$ (SSIM).
\begin{table}[]
\renewcommand{\arraystretch}{0.94}
    \centering
    \small
    \setlength{\tabcolsep}{2.52mm}
    \vspace{-0.90 em}
    \caption{Analysis of the number of LR and HR paths in Hi-scan.}
    \vspace{-0.60 em}
\begin{tabular}{cc|cc|ccc}
\toprule
LR & HR & FLOPs & Param. & NMSE   & SSIM   & PSNR  \\ \midrule
0       & 4       & 203G  & 3.19M & 0.0036 & 0.9722 & 39.14 \\
1       & 3       & 177G  & 2.79M & 0.0034 & 0.9731 & 39.37 \\
2       & 2       & 146G  & 2.31M & 0.0035 & 0.9725 & 39.23 \\
\rowcolor[HTML]{EFEFEF} 
3       & 1       & 117G  & 1.87M & 0.0034 & 0.9729 & 39.32 \\
4       & 0       & 98G  & 1.56M & 0.0044 & 0.8679 & 38.27 \\ \bottomrule
\end{tabular}
    \label{table7}
\vspace{-1.0 em}
\end{table}
\begin{table}[!t]
\renewcommand{\arraystretch}{0.94}
    \centering
    \small
    \setlength{\tabcolsep}{1.7mm}
    \caption{Efficiency comparison on CC359 dataset (AF=4) on RTX 3090.}
    \vspace{-0.60 em}
\begin{tabular}{l|cc|ccc}
\toprule
Method        & FLOPs & Param. & NMSE            & SSIM            & PSNR           \\ \midrule
MambaIR~\cite{r31}       & 190G  & 14.3M  & 0.0087          & 0.9506          & 35.13          \\
ReconFormer~\cite{r27}   & 342G  & 1.14M  & 0.0108          & 0.9297          & 34.16          \\
\rowcolor[HTML]{EFEFEF} 
\textbf{Ours} & 117G  & 1.87M  & \textbf{0.0034} & \textbf{0.9729} & \textbf{39.32} \\ \bottomrule
\end{tabular}
    \label{table8}
\vspace{-1.5 em}
\end{table}
\subsubsection{Effectiveness of Hierarchical Scanning}
To further demonstrate the effectiveness of our hierarchical scanning (Hi-scan), we present empirical results in Fig.\ref{Fig.8}. We compared the decay along scanning paths in SS2D~\cite{r11} and our method, focusing on the final token. As shown, SS2D suffers from significant long-range forgetting, with obvious decay along different paths. In contrast, our Hi-scan not only reduces long-range forgetting by shortening sequences via LR scanning but also preserves fine-grained details through the HR path, improving the model’s ability to capture global information.

We further conducted ablation studies on the number of LR and HR scanning paths. The results in Table \ref{table7} show that: (1) When all four scanning directions are HR (LR paths = $0$), the model achieves $39.14$ dB in reconstruction performance but requires high computational resources. (2) Increasing the number of LR scanning paths to $1$, $2$, and $3$ improves performance significantly while reducing computational load. This demonstrates the effectiveness of our Hi-scan in mitigating long-range forgetting. (3) When all four scanning directions are LR (LR paths = $4$), model performance drops sharply to $38.27$ PSNR, indicating that fully downsampled feature maps lose crucial fine-grained details. For the trade-off between performance and computational complexity, we set the number of LR paths $=3$ as the default setting.

\subsection{Further Analysis}
\subsubsection{Effective Receptive Fields}
We compared the effective receptive fields (ERFs) of several methods, which is visualized by computing the gradients of the output with respect to the input image. A wider distribution of dark areas demonstrates lager ERFs. As shown in Fig. \ref{Fig.9}: (1) CNN-based methods, including UNet, D5C5, DCRCN, and KIKI-Net, have limited ERFs, restricting their ability to capture global information. (2) Transformer-based methods, such as SwinMR, FMT-Net, and ReconFormer, achieve increased ERFs but still suffer from noticeable gradient decay from center to periphery. (3) Our DH-Mamba achieves the largest ERFs with linear computational complexity, enabling superior global exploration. This validates one of our main motivations that long-range dependency can be efficiently modeled via Mamba.

\subsubsection{Analysis of Training Efficiency}
The training efficiency comparison is reported in Table \ref{table8}. The recent ReconFormer employs a recurrent structure to maintain a few trainable parameters. However, its significantly higher computational complexity (FLOPs=$342$G) substantially increases both training difficulty and inference time. On the other hand, MambaIR addresses both spatial and channel redundancy and achieves a notable reduction in FLOPs. Nevertheless, MambaIR has a larger parameter count of $14.30$ M and  drop in performance. Compared to these methods, our DH-Mamba achieves significant performance improvements while maintaining both low complexity (FLOPs=$117$G) and a minimal number of trainable parameters ($1.87$M). This advantage is largely attributed to our Hi-scan, which mitigates the long-range forgetting and information redundancy while minimizing the computational complexity via down-sampling feature maps.
\section{Discussion}
Current MRI reconstruction methods face a trade-off between global receptive field and computational efficiency. To address this, we introduce Mamba for MRI reconstruction. We identify three challenges in applying Mamba into fast MRI: k-space distortion, long-range forgetting, and lack of local diversity. To overcome these issues, we propose DH-Mamba, which combines circular scanning design, dual-domain hierarchical processing, and local diversity enhancement. This approach enables efficient and effective MRI reconstruction, while also offering new insights for Mamba-based methods.


One limitation is that the Mamba's inherent causal modeling hinders precise MRI reconstruction. In standard Mamba, each token only depends on previous tokens in the scan sequence. The query pixel can only capture information from its predecessors and cannot perceive other pixels. Although multi-directional scanning alleviates this issue to some extent, it still cannot effectively capture information from arbitrary pixels that are strongly correlated with the query pixel. Additionally, it introduces extra computational burden. In future work, we will optimize Mamba for more effective scanning and modeling of MRI images, aiming for more accurate fast MRI reconstruction. We will also extend DH-Mamba to other recovery tasks, such as cardiac MRI.

\section{CONCLUSION}
This paper presents a Mamba-based model, named DH-Mamba, which pioneers the exploration of Mamba in dual-domain MRI reconstruction, providing an effective and efficient paradigm for this challenging application. Specifically, we design a k-space Mamba branch, which is customized for frequency domain global modeling. We introduce a hierarchical scanning technique that mitigates the long-range forgetting while minimizing the computational cost. We develop a local enhancement module that integrates local biases from CNN operators with spatial variation coordinates, enhancing the locality of feature representation and resulting in more accurate and reliable outcomes. Extensive experiments and analysis are conducted on CC359, fastMRI, and SKM-TEA datasets, validating the efficiency of our DH-Mamba to improve the performance while lower the computational costs.

\bibliographystyle{IEEEtran}
\bibliography{IEEEabrv,reference}
\vfill
\end{document}